\documentclass{aa}  
\newcommand{\msun}{$M_{\odot}$}
\usepackage{graphicx}
\usepackage{txfonts}
\usepackage{lscape} 
\begin{document}
   \title{Reaction rate uncertainties and the operation of the NeNa and MgAl chains during
HBB in intermediate-mass AGB stars\thanks{Table 13 is only available in electronic form
at the CDS via anonymous ftp to cdsarc.u-strasbg.fr (130.79.128.5)
or via http://cdsweb.u-strasbg.fr/cgi-bin/qcat?J/A+A/}
}

   \author{R.G. Izzard, 
          \inst{1}\fnmsep\thanks{The first two authors have contributed equally to this paper.}
          M. Lugaro,
          \inst{1,\star\star}
          A.I. Karakas,
          \inst{2}\fnmsep\thanks{Present address: Research School of Astronomy and Astrophysics,
                           Mt. Stromlo Observatory, Cotter Rd., Weston, ACT 2611, Australia,
                          \email{akarakas@mso.anu.edu.au}}
          C. Iliadis
          \inst{3}
          \and
          M. van Raai
          \inst{1}
          }

   \offprints{M. Lugaro}

   \institute{
        Sterrenkundig Instituut, University of Utrecht, 
        Postbus 80000 3508 TA Utrecht, The Netherlands\\
        \email{R.G.Izzard@phys.uu.nl}\\
        \email{M.Lugaro@phys.uu.nl}\\
        \email{raai@phys.uu.nl}
        \and
        Origins Institute, Department of Physics \& Astronomy,
         McMaster University, Hamilton ON, Canada\\
          \email{karakas@physics.mcmaster.ca}
         \and
         Department of Physics and Astronomy, University of North Carolina, 
         Chapel Hill, NC 27599-3255, USA; Triangle 
         Universities Nuclear Laboratory, P. O. Box 90308, Durham, 
         NC 27708-0308, USA\\
             \email{iliadis@unc.edu}
             }

   \date{Received 8 December, 2006}

 
\abstract
{We test the effect of proton-capture reaction rate uncertainties on the abundances of the Ne, Na, Mg 
and Al isotopes processed by the NeNa and MgAl chains during hot bottom burning (HBB) in 
asymptotic giant branch (AGB) stars of intermediate mass between 4 and 6 \msun\ and metallicities between 
$Z=0.0001$ and 0.02.} 
{We provide uncertainty ranges for the AGB stellar yields, for inclusion in galactic 
chemical evolution models, and indicate which reaction rates are most important and should be better 
determined.}
{We use a fast synthetic algorithm based on detailed AGB models. We run a large 
number of stellar models, varying one reaction per time for a very fine grid of values, as 
well as all reactions simultaneously.}
{We show that there are uncertainties in the yields of all the Ne, Na, Mg and Al isotopes 
due to uncertain proton-capture reaction rates. The most uncertain yields are those of 
$^{26}$Al and $^{23}$Na (variations of two orders of magnitude), $^{24}$Mg and $^{27}$Al (variations 
of more than one order of magnitude), $^{20}$Ne and $^{22}$Ne (variations between factors 2 
and 7). In order to obtain more reliable Ne, Na, Mg and Al yields from 
IM-AGB stars the rates that require more accurate determination are: 
$^{22}$Ne($p,\gamma$)$^{23}$Na, $^{23}$Na($p,\gamma$)$^{24}$Mg, $^{25}$Mg($p,\gamma$)$^{26}$Al, 
$^{26}$Mg($p,\gamma$)$^{27}$Al and $^{26}$Al($p,\gamma$)$^{27}$Si.} 
{Detailed galactic chemical evolution models should be constructed to address the impact of our uncertainty 
ranges on the observational constraints related to HBB nucleosynthesis, such as globular cluster chemical 
anomalies.}


\authorrunning{Izzard et al.}
\titlerunning{Reaction rate uncertainties and the NeNa, MgAl chains during HBB}

   \maketitle

\section{Introduction}

During the asymptotic giant branch (AGB) phase of stars (with initial masses between 
approximately 1 and 10 \msun) the 
abundances of several isotopes are modified via complex nucleosynthetic mechanisms. Hydrogen 
and helium burning occur alternately in shells in the deep layers of the star and mixing processes 
collectively known as the third dredge-up (TDU) carry the processed material to the 
stellar surface. Strong stellar winds eject the envelope of the star into the interstellar medium 
so that AGB stars contribute to the chemical evolution of galaxies. In AGB stars of masses higher 
than roughly 4 \msun\ (intermediate mass AGB stars, IM-AGB), H-burning occurs at the base of 
the convective H-rich envelope and its products are mixed to the surface of the star by convection. 
This process is called hot-bottom burning (HBB) and involves activation of the CNO, NeNa 
and MgAl cycles at temperatures between 60 and 100 million degrees. The activation of HBB in 
IM-AGB stars is validated by the fact that there appears to be an upper limit for the luminosity of 
carbon stars, in agreement with the fact that HBB would prevent the more massive AGB stars from becoming 
carbon rich (Boothroyd et al. \cite{boothroyd:93}). 

There are several types of applications for models of IM-AGB stars suffering HBB. Some 
direct observations of Li, C and the $^{12}$C/$^{13}$C ratios in IM-AGB stars are available, in particular for 
stars in the Magellanic Clouds (see e.g. Wood et al. \cite{wood:83}, and Plez et al. \cite{plez:93}), 
but also 
recently for our Galaxy (Garc\'ia-Herna\'ndez et al. \cite{garcia:06}). 
Stellar models are able to explain the fact that the majority of AGB stars of high luminosity are
O rich, as well as to reproduce the observed low $^{12}$C/$^{13}$C
ratios and high Li abundances (e.g. Boothroyd et al. \cite{boothroyd:93}, and Mazzitelli et al. 
\cite{mazzitelli:99}). More observational evidence of the occurrence of HBB comes from the fact that planetary 
nebulae of Type I, which are believed to come from massive AGBs, have enhanced He and N/O ratios, 
which can be explained by HBB in their AGB precursors (Peimbert \cite{peimbert:80}, Pottasch \& 
Bernard-Salas 
\cite{pottasch:06}, Stanghellini et al. \cite{stanghellini:06}).

Intermediate-mass AGB stars with HBB are candidates for pollution of globular cluster stars showing 
anomalies 
in O, Na, Mg, and Al (see review by Gratton et al. \cite{gratton:04}). Moreover, HBB combined 
with partial He burning and TDU make IM-AGB stars a site of production for at least part of the primary N 
observed at low metallicities (Spite et al. \cite{spite:05}). The effect of 
IM-AGB nucleosynthesis on the Mg isotopes has also been studied in relation to the apparent observed 
variation of the fine structure constant deduced from quasar absorption lines at redshift $<$ 2, which 
also depends on the abundance of the Mg isotopes (Ashenfelter et al. \cite{ashenfelter:04}, Fenner 
et al. \cite{fenner:05}). In this case, as well as in the study of abundance anomalies in globular 
clusters, AGB stellar yields have to be included into galactic chemical evolution (GCE) models 
in order to produce predictions to compare to the observable data (see e.g Fenner et al. 
\cite{fenner:04}). Finally, the origin of one meteoritic presolar spinel grain has 
been attributed to an IM-AGB star, providing the opportunity of studying massive AGBs 
using presolar grains (Lugaro et al. \cite{lugaro:06}). 

In summary, there are several applications of the study of HBB, however, IM-AGB stellar models and the 
resulting stellar yields have been calculated so far largely 
ignoring the effects of reaction-rate uncertainties. Ventura \& D'Antona (\cite{ventura:05a}) have 
produced two runs of one IM-AGB model using 
two sets of reaction rates: the old compilation of Caughlan \& Fowler (\cite{caughlan:88}) and the new 
NACRE compilation (Angulo et al \cite{angulo:99}). They found that uncertainties in the stellar 
structure physics, such as the mass-loss rate and the treatment of convection, produce 
errors in the resulting yields that are much larger than those produced by changing the rate 
compilation. However, comparing results for two different sets of rates does not exhaust the 
problem of testing reaction-rate uncertainties. Lugaro et al. (\cite{lugaro:04}) analyzed in detail the 
uncertainties of reaction rates in relation to the production of fluorine in AGB stars, and also 
discussed one IM-AGB model. Karakas et al. (\cite{karakas:06}) produced a  
detailed analysis of the effect in IM-AGB stars of the uncertainties of the rates 
for the $^{22}$Ne$+ \alpha$ reactions, which are responsible for the production of the 
heavy Mg isotopes during He burning. These authors presented new estimates for such rates, reducing 
the previous large errors of NACRE to much smaller ranges, and demonstrated that these rates do not 
constitute a source of uncertainties in IM-AGB models anymore. However, the precision with which the 
abundances of the Mg, as well as Ne, Na and Al, isotopes in IM-AGB stars can be predicted is still undermined by 
uncertainties in the proton capture reaction involved in the NeNa and MgAl chains. Ventura \& D'Antona 
(\cite{ventura:06}) considered in particular the effect of the uncertainties in the rates the produce and destroy 
$^{23}$Na in their $M=5$ \msun\ and $Z=0.001$ model.

With the present work we attempt to fill the gap in the analysis of uncertainties connected to the 
yields from IM-AGB stars. Our analysis has two main motivations: we want to provide uncertainties for 
the stellar yields, for inclusion in GCE models, and we want to indicate which reaction rates require 
better determination. Our work is targeted specifically at the NeNa and MgAl chains during HBB. The 
Ne, Na, Mg and Al isotopes are of interest in all the applications of IM-AGB star nucleosynthesis 
listed above and our calculation tools are most suitable to this problem: we can treat these 
proton-capture processes in an analytical way and produce a very large number of stars in a 
reasonable time. It is not possible to perform the same analysis with our method for the CNO cycle 
because changing the rates in the CNO cycle would affect the stellar structure, i.e. the temperature 
at the base of the convective envelope and the convection timescale, which are not calculated in our 
synthetic post-processing but taken from the detailed AGB models.

The present study has been inspired by two previous works aimed at systematically testing the effect of 
reaction rate uncertainties: the paper on nova nucleosynthesis by Iliadis et al. 
(\cite{iliadis:02}) and the paper on the oxygen isotopic ratios in red giant stars by Stoesz \& Herwig 
(\cite{stoesz:03}). While Iliadis et al. (\cite{iliadis:02}) could only test one rate at a time, and 
Stoesz \& Herwig (\cite{stoesz:03}) used a Monte Carlo approach to run the grid of models varying the 
reaction rates, the method we use is so fast that we can run a large number (typically 10$^4$) of 
stellar models varying one reaction at a time for a very fine grid of values, as 
well as 
all reactions simultaneously. Still, our models are based on 
fully-evolved stellar structure models.

The paper is structured as follows: in Sec. 2 we describe our methods and models in detail, in Sec. 3 we 
list the reaction rates and the uncertainties we have employed. In Sec. 4 we present the results. In Sec. 5 
we summarize the results, and present our conclusions. 

As a final introductory comment we observe that it is not in our aims to discuss any of the major 
uncertainties that are still related to the stellar structure of IM-AGB models, such as the mass-loss law, 
the treatment of convection, and the determination of convective borders. Discussions of these can be found 
for example in Herwig (\cite{herwig:05}), Ventura \& D'Antona (\cite{ventura:05a,ventura:05b}), Karakas et al. 
(\cite{karakas:06}), Lugaro et al. (\cite{lugaro:06}). These uncertainties are typically large and difficult to 
estimate, calling for more work to be done on the physics of AGB models.

\section{Methods and models}

We use the single and binary synthetic nucleosynthesis model of Izzard et al. (\cite{izzard:06}, 
hereafter: the {\it synthetic} models), 
where third dredge-up is followed according to the prescriptions of Karakas et al. 
(\cite{karakas:02}). The hot bottom burning model is described by Izzard et al. (\cite{izzard:04}) 
and updated by Izzard et al. (\cite{izzard:06}). It approximates HBB in AGB stars by replacing the 
many burning and mixing cycles of a detailed stellar evolutionary calculation with a single burning 
and mixing event at each pulse. The third dredge-up, the fraction of the stellar envelope exposed to 
HBB, and the burning timescale are calibrated to the detailed stellar evolution models of Karakas et 
al. (\cite{karakas:02}). The maximum temperature at the base of the convective envelope ($T_{\rm 
bce}^{\rm max}$) and the total amount of third dredge-up ($M_{\rm TDU}^{\rm tot}$) obtained from the 
detailed calculations for all the stellar models considered here are presented in Table 
\ref{tab:models}. The CNO, NeNa and MgAl cycles are followed by analytic solution of the appropriate 
differential rate equations, which means our synthetic model is both extremely fast and accurate. We 
follow both the ground and metastable state of $^{26}$Al, but it turned out that introducing the 
metastable state does not alter our results significantly.

The contribution to luminosity or opacity, and hence stellar structure, of the NeNa and MgAl cycles is 
negligible, so we make the assumption that we can vary the rates of nuclear reactions involving these 
species without changing the physical parameters of the star such as interpulse periods and the amount of 
third dredge-up. We could not, for example, change the CNO cycle reaction rates, as these alter the 
interpulse period and the amount of third dredge-up (Herwig \& Austin \cite{herwig:04}).

\begin{table}
\begin{minipage}[t]{\columnwidth}
\caption{Maximum temperature at the base of the convective envelope ($T_{\rm bce}^{\rm max}$), and total 
third dredge-up mass ($M_{\rm TDU}^{\rm tot}$) extracted from the detailed calculations for all the models 
discussed here.}             
\label{tab:models}      
\centering     
\renewcommand{\footnoterule}{}  
\begin{tabular}{ccc}    
\hline\hline   
Model($M$,$Z$) & $T_{\rm bce}^{\rm max}$/10$^7$ K & $M_{\rm TDU}^{\rm tot}$ (\msun) \\  
\hline                        
5,0.02 & 6.26 & 0.050 \\
6,0.02 & 8.26 & 0.058 \\
5,0.008 & 8.03 & 0.180 \\
6,0.008 & 8.90 & 0.126 \\
5,0.004 & 8.39 & 0.225 \\
6,0.004 & 9.40 & 0.151 \\
4,0.0001 & 8.22 & 0.302 \\
5,0.0001 & 9.10 & 0.320 \\
6,0.0001 & 10.3 & 0.119 \\
\hline  
\end{tabular}
\end{minipage}
\end{table}

The stellar yields for the Ne, Na, Mg, and Al isotopes calculated by the synthetic models by 
setting all the reaction rates to their recommended values (see details in Sec. \ref{sec:rates} and 
Table \ref{tab:rates}) are 
listed in Table \ref{tab:control}. These yields are defined as the total mass ejected of each isotope 
in solar masses and represent our control values. In the same table we also present the yields from 
the models of Karakas et al. (\cite{karakas:02}, hereafter: the {\it detailed} models) calculated by 
setting all the reaction rates to the same recommended values used in the synthetic models. 
Note that initial compositions for the models with metallicity lower than solar are always taken to 
be scaled 
solar. IM-AGB stars are important 
producers of $^{22}$Ne, $^{25}$Mg, $^{26}$Mg and $^{26}$Al. The galactic production of 
$\alpha$-nuclei $^{20}$Ne and $^{24}$Mg and that of $^{27}$Al is instead dominated by supernova 
nucleosynthesis, even if IM-AGB nucleosynthesis can affect the abundances of $^{24}$Mg, which can be 
heavily destroyed in low-metallicity models and of $^{27}$Al, which can be slightly produced.

The ratio between the detailed and the synthetic models are typically within a factor of two, except 
for the isotopes of lowest abundance: $^{21}$Ne (ratios up to 4) and $^{26}$Al (ratios up to 5.4) and 
for the $M=6$ \msun\ and $Z=0.0001$ model. These differences come up for a number of reasons:

\begin{enumerate}

\item{In the synthetic models the AGB evolution is followed right to the end, i.e. until the envelope is 
completely 
lost, while in the detailed models the AGB evolution is followed up to a point when the code does not converge 
anymore or when an enormous amount of models has been generated. To calculate the yields it is 
assumed 
that the rest of the envelope is ejected with the final computed 
composition. Hence, more thermal pulses and a longer time for HBB is considered in the calculation of the synthetic 
yields. In particular, this explains the differences obtained for the $M=6$ \msun\ and $Z=0.0001$ model, which are 
larger than for the other models: in fact, with the detailed models we calculated 106 thermal pulses and then the 
computation was stopped, however, the total mass of the star was still 5.96 \msun\ at this point of the 
evolution.\footnote{Note that the $M=6$ \msun\ and $Z=0.0001$ model has 691,973 evolutionary steps, and it 
takes almost two weeks to compute the detailed nucleosynthesis on a 
AMD athlon 3500+ 64bit, ASUS A8V deluxe machine. The $M=5$ \msun\ and 
$Z=0.0001$ model, for which we have calculated 136 thermal pulses with a final envelope mass of 0.76 \msun, has 
1,278,389 evolution models and it takes four weeks to compute the detailed nucleosynthesis. It is clearly not 
computationally feasible to get to the end of the evolution for the $M=6$ \msun\ and $Z=0.0001$ model with 
the detailed calculations.}}

\item{The effect of 
the second dredge-up on the elements considered here is not accounted for in the synthetic models.} 

\item{The dredge-up of the thin layer of H-burning ashes in the intershell not engulfed in the thermal 
pulses is 
accounted for in the detailed but not in the synthetic models;}
 
\item{The $M=4$ and $M=5$ \msun\ Z=0.0001 models present degenerate pulses (Frost et al. 
\cite{frost:98}), which 
are handled properly by the detailed models but are not included in the synthetic model.}

\item{Finally, of course, the synthetic algorithm is by its nature approximate.} 

\end{enumerate}

They relatively small differences between our detailed and synthetic yields are well within stellar 
model uncertainties and, in many cases, also within reaction rate uncertainties (see Sec. 
\ref{sec:results}). In any case, we have tested for the $M=6$ \msun\ Z=0.02 and Z=0.004 models that 
the range in the yields derived from the synthetic models by varying the reaction rates within their 
uncertainties is the same 
as derived from the detailed models for the important cases of the upper limits of the 
$^{25}$Mg$(p,\gamma$)$^{26}$Al and $^{26}$Al$(p,\gamma$)$^{27}$Si reaction rates. These rates affect 
the yield of $^{26}$Al, which is one of the isotopes where the detailed and the synthetic models 
disagree most. We found that the ranges obtained by the detailed models are very close to those 
obtained by the synthetic models. More details are given in Sec.~\ref{sec:results}. For the other 
isotope where the detailed and the synthetic models disagree most, $^{21}$Ne, we did not do the same 
exercise because the abundance is too low to make this isotope unimportant, and there are no 
uncertainties on it derived from the reaction rates (see Sec.~\ref{sec:results}).

\begin{table*}
\caption{Control values for the yields computed with the synthetic rapid code (first line for each model,
e.g. 7.118e-03 stands for $7.118 \times 10^{-3}$) and with the 
detailed models (second line for each model, in italics). The ratio between the two is presented in 
the third line for each model.}
\label{tab:control}      
\centering
\begin{tabular}{cccccccccc}    
\hline\hline   
mass,metallicity & $^{20}$Ne & $^{21}$Ne & $^{22}$Ne & $^{23}$Na & $^{24}$Mg & $^{25}$Mg & 
$^{26}$Mg & $^{26}$Al & 
$^{27}$Al \\  
\hline    
5,0.02 & 7.118e-03 & 1.223e-05 & 1.771e-03 & 2.234e-04 & 2.231e-03 & 4.661e-04 & 7.541e-04 & 8.252e-07 & 2.646e-04 \\
       & {\it 6.681e-03} & {\it 1.965e-05} & {\it 1.390e-03} & {\it 2.291e-04} & {\it 2.107e-03} & {\it 3.431e-04} & 
{\it 4.388e-04} & {\it 4.966e-07} & {\it 2.477e-04} \\
       & 1.06 & 0.62 & 1.27 & 0.98 & 1.06 & 1.36 & 1.72 & 1.66 & 1.07 \\ 
6,0.02 & 8.770e-03 & 2.237e-06 & 2.047e-03 & 2.711e-04 & 2.690e-03 & 6.990e-04 & 1.040e-03 & 1.172e-05 & 3.373e-04 \\ 
       & {\it 8.211e-03} & {\it 1.067e-06} & {\it 1.534e-03} & {\it 3.012e-04} & {\it 2.557e-03} & {\it 4.817e-04} & 
{\it 6.383e-04} & {\it 4.096e-06} & {\it 3.128e-04} \\
       & 1.07 & 2.09 & 1.33 & 0.90 & 1.05 & 1.45 & 1.63 & 2.86 & 1.08 \\     
5,0.008 & 2.974e-03 & 1.463e-06 & 2.175e-03 & 8.740e-05 & 8.445e-04 & 6.275e-04 & 8.250e-04 & 2.085e-05 & 1.488e-04 \\ 
        & {\it 2.678e-03} & {\it 5.437e-07} & {\it 2.379e-03} & {\it 1.202e-04} & {\it 7.861e-04} & {\it 4.307e-04} & 
{\it 6.922e-04} & {\it 8.961e-06} & {\it 1.284e-04} \\
        & 1.11 & 2.69 & 0.91 & 0.73 & 1.07 & 1.46 & 1.19 & 2.32 & 1.16 \\  
6,0.008 & 3.604e-03 & 1.213e-06 & 1.852e-03 & 7.352e-05 & 5.878e-04 & 1.044e-03 & 9.582e-04 & 1.037e-04 & 1.855e-04 \\ 
        & {\it 3.289e-03} & {\it 4.463e-07} & {\it 1.233e-03} & {\it 1.072e-04} & {\it 6.711e-04} & {\it 6.929e-04} & 
{\it 6.389e-04} & {\it 2.527e-05} & {\it 1.430e-04} \\
        & 1.10 & 2.72 & 1.50 & 0.69 & 0.88 & 1.51 & 1.40 & 4.10 & 1.30 \\
5,0.004 & 1.555e-03 & 1.569e-06 & 2.576e-03 & 4.908e-05 & 3.544e-04 & 7.456e-04 & 1.184e-03 & 5.012e-05 & 1.007e-04 \\ 
        & {\it 1.360e-03} & {\it 5.780e-07} & {\it 2.371e-03} & {\it 9.156e-05} & {\it 3.390e-04} & {\it 4.934e-04} & 
{\it 9.944e-04} & {\it 1.469e-05} & {\it 8.363e-05} \\
        & 1.14 & 2.71 & 1.09 & 0.54 & 1.04 & 1.51 & 1.19 & 3.41 & 1.20 \\ 
6,0.004 & 1.830e-03 & 1.108e-06 & 1.374e-03 & 2.553e-05 & 5.730e-05 & 8.156e-04 & 8.629e-04 & 1.652e-04 & 1.526e-04 \\ 
        & {\it 1.664e-03} & {\it 3.338e-07} & {\it 9.963e-04} & {\it 4.506e-05} & {\it 6.782e-05} & {\it 6.999e-04} & 
{\it 7.182e-04} & {\it 5.223e-05} & {\it 9.631e-05} \\ 
        & 1.10 & 3.32 & 1.38 & 0.57 & 0.84 & 1.16 & 1.20 & 3.16 & 1.58 \\
4,0.0001 & 1.741e-04 & 1.563e-06 & 4.047e-03 & 3.254e-05 & 1.766e-05 & 4.121e-04 & 1.684e-03 & 1.478e-04 & 4.297e-05 \\ 
         & {\it 2.156e-04} & {\it 8.663e-07} & {\it 3.697e-03} & {\it 1.315e-04} & {\it 3.578e-05} & {\it 4.864e-04} & 
{\it 1.664e-03} & {\it 2.747e-05} & {\it 1.214e-04} \\
         & 0.81 & 1.80 & 1.09 & 0.25 & 0.49 & 0.85 & 1.01 & 5.38 & 0.35 \\ 
5,0.0001 & 2.436e-04 & 1.515e-06 & 3.128e-03 & 3.703e-05 & 3.772e-05 & 5.271e-04 & 2.107e-03 & 5.038e-05 & 1.489e-04 \\ 
         & {\it 1.642e-04} & {\it 3.878e-07} & {\it 3.088e-03} & {\it 9.655e-05} & {\it 3.010e-05} & {\it 5.799e-04} 
& {\it 2.110e-03} & {\it 2.333e-05} & {\it 1.077e-04} \\
         & 1.48 & 3.91 & 1.01 & 0.38 & 1.25 & 0.91 & 1.00 & 2.16 & 1.38 \\
6,0.0001 & 2.146e-04 & 8.538e-07 & 1.888e-03 & 2.474e-05 & 5.588e-06 & 3.674e-04 & 1.373e-03 & 4.399e-05 & 1.716e-04 \\ 
         & {\it 6.973e-05} & {\it 1.461e-08} & {\it 3.423e-04} & {\it 8.766e-06} & {\it 1.268e-06} & {\it 8.058e-05} & 
{\it 2.667e-04} & {\it 6.145e-06} & {\it 3.211e-05} \\ 
         & 3.08 & 58.4 & 5.51 & 2.82 & 4.40 & 4.56 & 5.15 & 7.16 & 5.34 \\
\hline  
\end{tabular}
\end{table*}

\section{The choice of reaction rates and their uncertainties}
\label{sec:rates}

\begin{table*}
\caption{References and uncertainties (in the range T = 70 - 100 $\times 10^6$ K) for the considered 
reaction rates.} 
\label{tab:rates}      
\begin{center}
\renewcommand{\footnoterule}{}  
\begin{tabular}{c c c c c}        
\hline\hline                 
Rate & Energy$^a$ (keV) & Reference & Uncertainty & Chosen Uncertainty \\    
\hline                        
$^{20}$Ne($p,\gamma$)$^{21}$Na & 67 - 157 & Iliadis et al. (\cite{iliadis:01}) & /2, $\times$1.5 & /2, 
$\times$1.5 \\ 
 & & = NACRE & \\
$^{21}$Ne($p,\gamma$)$^{22}$Na & 67 - 157 & Iliadis et al. (\cite{iliadis:01}) & /1.25, $\times$1.20 & /1.25, 
$\times$1.20 \\ 
$^{22}$Ne($p,\gamma$)$^{23}$Na & 67 - 157 & Iliadis et al. (\cite{iliadis:01}) & /1.43 to /2, $\times$982 to 
$\times$1888 & /2, $\times$2000 \\ 
 & & = Hale et al. (\cite{hale:02}) & \\
$^{23}$Na($p,\gamma$)$^{24}$Mg & 73 - 166 & Rowland et al. (\cite{rowland:04}) & /5 to /40, $\times$7.8 to 
$\times$9.8 & /40, $\times$10 \\ 
$^{23}$Na($p,\alpha$)$^{20}$Ne & 73 - 166 & Rowland et al. (\cite{rowland:04}) & /1.3, $\times$1.3 & /1.3, 
$\times$1.3 \\ 
$^{24}$Mg($p,\gamma$)$^{25}$Al & 78 - 175 & Iliadis et al. (\cite{iliadis:01}) & /1.2, $\times$1.2 & /1.2, 
$\times$1.2 \\ 
 & & = Powell et al. (\cite{powell:99}) & \\
$^{25}$Mg($p,\gamma$)$^{26}$Al & 78 - 175 & Iliadis et al. (\cite{iliadis:01}) & /2, $\times$ 1.5 & /2, $\times$ 
1.5 \\ 
$^{26}$Mg$(p,\gamma$)$^{27}$Al & 78 - 175 & Iliadis et al. (\cite{iliadis:01}) & /1.4 to /4.2, $\times$4.1 to 
$\times$8.9 & /4, $\times$10 \\ 
$^{26}$Al$_{ground}(p,\gamma$)$^{27}$Si & 83 - 184 & Iliadis et al. (\cite{iliadis:01}) & /1.25 to /2, $\times$11. 
to $\times$572 & /2, $\times$600 \\ 
$^{27}$Al$(p,\gamma$)$^{28}$Si & 83 - 184 & Iliadis et al. (\cite{iliadis:01}) & /1.25, $\times$1.2 to $\times$2.8 
& /1.25, $\times$3 \\ 
\hline                                   
\end{tabular}
\end{center}
$^a$ Effective stellar energy window for the reaction, calculated as ($E_0 - \Delta/2)_{70 {\rm MK}}$ - ($E_0 
+ \Delta/2)_{100 {\rm MK}}$, where $E_0$ is the location of the Gamow peak and $\Delta$ is its $1/e$ width.
\end{table*}

The rate references, stellar energy windows and uncertainties are presented in Table \ref{tab:rates}. 
The bulk of the rates and their uncertainties come from the compilation 
of Iliadis et al. (\cite{iliadis:01}), except for the $^{23}$Na$+p$ rates which come from the more recent 
work of Rowland et al (\cite{rowland:04}). The 
uncertainty ranges and stellar energy windows are for the range of temperatures relevant to the activation of 
the 
NeNa and MgAl chains during HBB: 70 - 100 $\times 10^6$~K. The 
uncertainties are expressed as multiplication and division factors of the recommended rates to obtain the upper 
and the lower limits of the rates, respectively. The uncertainty ranges of Column 4 describe in details how the 
uncertainty varies in the given range of temperature. The actual uncertainty ranges used in 
our calculations have been derived as the maximum values from Column 4 and are listed in Column 5.
For the $^{20}$Ne($p,\gamma$)$^{21}$Na, 
$^{21}$Ne($p,\gamma$)$^{22}$Na, $^{23}$Na($p,\alpha$)$^{20}$Ne, $^{24}$Mg($p,\gamma$)$^{25}$Al, 
$^{25}$Mg$(p,\gamma$)$^{26}$Al reactions the uncertainty factors are approximately constant in the HBB 
temperature range, thus taking constant factors is a good description. We have checked that this is 
true by 
running the detailed $M=6$ \msun, Z=0.02 and Z=0.004 models using a more accurate description of the 
upper limit of the 
$^{25}$Mg$(p,\gamma$)$^{26}$Al rate and obtained the same results as described in Sec.~\ref{sec:results} 
within 7\%. For the reactions for which the uncertainty factors are not approximately constant we discuss at the end 
of Sec.~\ref{sec:examples} if our choice of the uncertainty range make the results less reliable.

The largest reaction rate uncertainties, up to three orders of magnitude, are associated with the 
$^{22}$Ne($p,\gamma$)$^{23}$Na and $^{26}$Al$(p,\gamma$)$^{27}$Si rates. They are caused by contributions from 
as yet unobserved low-energy resonances. We expect these rate uncertainties to have the largest impact on the 
yields. Other well determined reaction rates, such as the $^{24}$Mg($p,\gamma$)$^{25}$Al rate, are not expected 
to have a large effect on the stellar yields.

To each value of the rate in between the lower and the upper limits a probability has to be assigned. 
Ideally, having available all the information on nuclear properties to calculate the rates, it would be 
appropriate to use a log-normal distribution (see Thompson \& Iliadis \cite{thompson:99}). However, there 
are two problems with this. First, for many rates the uncertainty ranges are very asymmetrical even on a 
logarithmic scale. Second, we already pointed out that the large rate uncertainties at relatively low 
temperatures are caused by as yet unobserved low-energy resonances. The rates are in such cases obtained in 
the following way (Angulo et al. \cite{angulo:99}; Iliadis et al. \cite{iliadis:01}): (i) a maximum 
possible contribution is estimated - based on theoretical models or measured upper limits for the resonance 
strengths - for all threshold states; the inclusion of this contribution provides the upper total rate 
limit; (ii) disregarding any contributions from the threshold states provides the lower total rate limit; 
and (iii) the recommended total rates are then arrived at by multiplying the upper limit contributions of 
the threshold states by an (arbitrary) factor of 0.1. It should be clear from this description that there 
is no straightforward manner for representing the reaction rate errors by a meaningful probability 
distribution function. For these reasons, we have decided to assign the same probability to each value of 
the rates between the upper and the lower limits, i.e. to use a flat probability distribution. The choice 
of a probability distribution does not influence our resulting ranges of uncertainty in the yields, which 
remain a strong result of our work. However, it defines how the yield values are distributed, and hence, 
for example, which is the most probable value of the distribution associated with each yield. In Sec. 
\ref{sec:examples} we present two examples of the probability distribution of the yields that we obtain, 
keeping in mind that, while assigning a probability distribution to the rates is still difficult, these 
examples should be only considered as test exercises.

\section{Results}
\label{sec:results}

Tables \ref{tab:ne20} to \ref{tab:al27} present the range of uncertainties we obtain for the Ne, Na, Mg and 
Al isotopes when varying the reaction rates within their uncertainty
ranges. Only variations of more than 10\% are listed. In each table we present yield variations relative to the 
control value (see Table~\ref{tab:control}) as functions of the stellar model (Column 1: mass in \msun\ and 
metallicity) and of the reaction rates (headers). The last column gives the range of uncertainties we obtain when 
we vary all the reaction rates simultaneously in all possible combinations of lower and upper limits. 
The uncertainties in the yields obtained when varying simultaneously all the reaction rates are larger than 
those obtained by vary each single rate since in this case the uncertainties from all rates are applied, 
however, they do not always correspond simply to multiplying the factors 
obtained by varying each single rate. This indicates the complex interplay of the reactions involved in the NeNa and 
MgAl chains, and points out the importance of computing models using all possible combinations of rates. 

\begin{table}
\begin{minipage}[t]{\columnwidth}
\caption{Multiplication factors for the $^{20}$Ne yields.}             
\label{tab:ne20}      
\centering     
\renewcommand{\footnoterule}{}  
\begin{tabular}{ccc}    
\hline\hline   
Model & $^{22}$Ne($p,\gamma$)$^{23}$Na & All reactions \\  
\hline                        
5,0.004 & 1.0 - 1.11 & 0.99 - 1.15 \\
6,0.004 & 1.0 - 1.25 & 0.98 - 1.29 \\
4,0.0001 & 1.0 - 1.46 & 1.0 - 1.60 \\
5,0.0001 & 0.95 - 5.50 & 0.93 - 6.20 \\
6,0.0001 & 0.93 - 4.34 & 0.90 - 4.75 \\
\hline  
\end{tabular}
\end{minipage}
\end{table}

\begin{table}
\begin{minipage}[t]{\columnwidth}
\caption{Multiplication factors for the $^{21}$Ne yields.}             
\label{tab:ne21}      
\centering     
\renewcommand{\footnoterule}{}  
\begin{tabular}{cc}    
\hline\hline   
Model & All reactions \\  
\hline                        
6,0.0001 & 1.0 - 1.11 \\
\hline  
\end{tabular}
\end{minipage}
\end{table}

\begin{table}
\begin{minipage}[t]{\columnwidth}
\caption{Multiplication factors for the $^{22}$Ne yields.}             
\label{tab:ne22}      
\centering     
\renewcommand{\footnoterule}{}  
\begin{tabular}{ccc}    
\hline\hline   
Model & $^{22}$Ne($p,\gamma$)$^{23}$Na & All reactions \\  
\hline     
5,0.02 & 0.83 - 1.0 & 0.83 - 1.0 \\
6,0.02 & 0.33 - 1.0 & 0.33 - 1.0 \\
5,0.008 & 0.20 - 1.0 & 0.20 - 1.0 \\                   
6,0.008 & 0.18 - 1.0 & 0.18 - 1.01 \\                   
5,0.004 & 0.18 - 1.0 & 0.18 - 1.0  \\                   
6,0.004 & 0.17 - 1.01 & 0.17 - 1.02 \\
4,0.0001 & 0.17 - 1.0 & 0.17 - 1.0 \\
5,0.0001 & 0.14 - 1.01 & 0.14 - 1.01 \\
6,0.0001 & 0.17 - 1.01 & 0.17 - 1.01 \\
\hline  
\end{tabular}
\end{minipage}
\end{table}

\begin{table*}
\begin{minipage}[t]{\columnwidth}
\caption{Multiplication factors for the $^{23}$Na yields.}             
\label{tab:na23}      
\centering     
\renewcommand{\footnoterule}{}  
\begin{tabular}{ccccc}    
\hline\hline   
Model & $^{22}$Ne($p,\gamma$)$^{23}$Na & $^{23}$Na($p,\gamma$)$^{24}$Mg & 
$^{23}$Na($p,\alpha$)$^{20}$Ne & All reactions \\  
\hline     
5,0.02 & 1. - 2.41 & & & 1. - 2.41 \\
6,0.02 & 1. - 6.21 & & & 0.97 - 6.25 \\
5,0.008 & 0.95 - 21.3 & & & 0.91 - 21.5 \\                   
6,0.008 & 0.89 - 17.9 & 0.86 - 1.02 & 0.90 - 1.09 & 0.68 - 19.3 \\                   
5,0.004 & 0.80 - 41.6 & & & 0.70 - 42.8 \\                   
6,0.004 & 0.67 - 26.7 & 0.80 - 1.03 & 0.83 - 1.18 & 0.43 - 30.9 \\
4,0.0001 & 0.62 - 106.3 & & & 0.61 - 107.2 \\
5,0.0001 & 0.53 - 41.3 & 0.86 - 1.02 & 0.88 - 1.12 & 0.41 - 46.8 \\
6,0.0001 & 0.56 - 32.6 & 0.86 - 1.02 & 0.87 - 1.13 & 0.47 - 36.5 \\
\hline  
\end{tabular}
\end{minipage}
\end{table*}

\begin{table*}
\begin{minipage}[t]{\columnwidth}
\caption{Multiplication factors for the $^{24}$Mg yields.}             
\label{tab:mg24}      
\centering     
\renewcommand{\footnoterule}{}  
\begin{tabular}{ccccc}    
\hline\hline   
Model & $^{22}$Ne($p,\gamma$)$^{23}$Na & $^{23}$Na($p,\gamma$)$^{24}$Mg & 
$^{24}$Mg($p,\gamma$)$^{25}$Al & All reactions \\  
\hline     
6,0.008 & & & 0.89 - 1.10 & 0.89 - 1.34 \\                   
5,0.004 & & & & 0.96 - 1.37 \\                   
6,0.004 & 0.99 - 1.24 &  & 0.73 - 1.40 & 0.72 - 4.09 \\
4,0.0001 & 1. - 1.41 &  & & 0.99 - 5.11 \\
5,0.0001 & 0.94 - 5.27 & 0.88 - 1.94 & & 0.81 - 48.0 \\
6,0.0001 & 0.98 - 2.20 & 0.95 - 1.37 & & 0.91 - 14.3 \\
\hline  
\end{tabular}
\end{minipage}
\end{table*}

\begin{table}
\begin{minipage}[t]{\columnwidth}
\caption{Multiplication factors for the $^{25}$Mg yields.}  
\label{tab:mg25}      
\centering     
\renewcommand{\footnoterule}{}  
\begin{tabular}{ccc}    
\hline\hline   
Model & $^{25}$Mg($p,\gamma$)$^{26}$Al & All reactions \\  
\hline     
6,0.008 & & 0.90 - 1.16 \\
6,0.004 & 0.90 - 1.12 & 0.90 - 1.25 \\
5,0.0001 & & 0.94 - 1.67 \\
6,0.0001 & & 0.93 - 1.66 \\
\hline  
\end{tabular}
\end{minipage}
\end{table}

\begin{table}
\begin{minipage}[t]{\columnwidth}
\caption{Multiplication factors for the $^{26}$Mg yields.}             
\label{tab:mg26}      
\centering     
\renewcommand{\footnoterule}{}  
\begin{tabular}{ccc}    
\hline\hline   
Model & $^{26}$Mg($p,\gamma$)$^{27}$Al & All reactions \\  
\hline     
6,0.008 & 0.80 - 1.02 & 0.78 - 1.03 \\
5,0.004 & 0.88 - 1.01 & 0.87 - 1.01 \\
6,0.004 & 0.72 - 1.04 & 0.69 - 1.06 \\
5,0.0001 & 0.73 - 1.03 & 0.72 - 1.05 \\
6,0.0001 & 0.65 - 1.06 & 0.65 - 1.07 \\
\hline  
\end{tabular}
\end{minipage}
\end{table}

\begin{table}
\begin{minipage}[t]{\columnwidth}
\caption{Multiplication factors for the $^{26}$Al yields. In italics are the 
variations derived by running the corresponding detailed models.}             
\label{tab:al26}      
\centering     
\renewcommand{\footnoterule}{}  
\begin{tabular}{cccc}    
\hline\hline   
Model & $^{25}$Mg($p,\gamma$)$^{26}$Al & 
$^{26}$Al($p,\gamma$)$^{27}$Si & All reactions \\  
\hline
5,0.02 & 0.81 - 1.19 & & 0.80 - 1.19 \\
6,0.02 & 0.56 - 1.44 ({\it 1.34}) & 0.46 ({\it 0.43}) - 1.0 & 0.27 - 1.46 \\                        
5,0.008 & 0.52 - 1.47 & 0.36 - 1.0 & 0.18 - 1.48 \\
6,0.008 & 0.52 - 1.44 & 0.07 - 1.02 & 0.04 - 1.61 \\
5,0.004 & 0.52 - 1.44 & 0.14 - 1.01 & 0.07 - 1.51 \\
6,0.004 & 0.55 - 1.38 ({\it 1.35}) & 0.02 ({\it 0.03}) - 1.07 & 0.01 - 1.70 \\
4,0.0001 & 0.51 - 1.46 & 0.38 - 1.0 & 0.20 - 1.47 \\
5,0.0001 & 0.53 - 1.41 & 0.03 - 1.07 & 0.02 - 2.70 \\
6,0.0001 & 0.56 - 1.36 & 0.03 - 1.16 & 0.02 - 3.03 \\
\hline  
\end{tabular}
\end{minipage}
\end{table}

\begin{table*}
\begin{minipage}[t]{\columnwidth}
\caption{Multiplication factors for the $^{27}$Al yields.
In italics are the variations derived by running the corresponding detailed models.}             
\label{tab:al27}      
\centering     
\renewcommand{\footnoterule}{}  
\begin{tabular}{cccc}    
\hline\hline   
Model & $^{26}$Mg($p,\gamma$)$^{27}$Al & 
$^{26}$Al($p,\gamma$)$^{27}$Si & All reactions \\  
\hline
6,0.02 & 0.99 - 1.12 & & 0.99 - 1.15 \\
5,0.008 & 0.97 - 1.36 & 1.0 - 1.12 & 0.97 - 1.54 \\
6,0.008 & 0.88 - 2.09 & 0.99 - 1.65 & 0.87 - 3.14 \\
5,0.004 & 0.86 - 2.46 & 1.0 - 1.58 & 0.85 - 3.32 \\
6,0.004 & 0.78 - 2.63 & 0.91 - 2.29 ({\it 1.59$^{a}$}) & 0.62 - 4.78 \\
4,0.0001 & 0.79 - 3.38 & 1.0 - 1.31 & 0.78 - 3.85 \\
5,0.0001 & 0.48 - 4.97 & 0.97 - 1.43 & 0.43 - 6.15 \\
6,0.0001 & 0.47 - 3.91 & 0.95 - 1.29 & 0.39 - 4.88 \\
\hline  
\end{tabular}
\end{minipage}

$^{a}$This result from the detailed model is closer to what expected by a comparison to 
models of other masses and metallicities. We do not yet have a ready explanation for the 
44\% higher range shown by the synthetic model.  
\end{table*}

The isotope least affected by reaction-rate uncertainties is $^{21}$Ne, which is also typically destroyed 
by HBB and thus has very low stellar yields. All the other isotopes show important variations in their 
yields, which of course increase in magnitude with increasing the HBB efficiency. This is mostly 
determined by the temperature at the base of the 
convective envelope, which increases with decreasing metallicity and increasing stellar mass (Table 
\ref{tab:models}). 
Thus, the rate uncertainties typically have a larger impact on models of higher masses and lower 
metallicities. On the other hand, the TDU mass is also an important parameter in determining the HBB 
efficiency, as it feeds fresh $^{22}$Ne, $^{25}$Mg and $^{26}$Mg nuclei from the He intershell into the 
envelope to be burned by the NeNa and MgAl chains. The TDU mass increases with decreasing the metallicity, 
but, contrary to the HBB temperature, it typically decreases with increasing the stellar mass in IM-AGB models 
(Table \ref{tab:models}). The 
combined effect of TDU mass and HBB temperature eventually determines the HBB efficiency and hence the 
effect of the rate uncertainties. For example, the uncertainty range of $^{23}$Na increases with decreasing 
the metallicity (Table \ref{tab:na23}), however, it decreases with the stellar mass at any given 
metallicity lower than solar, 
because less TDU means less $^{22}$Ne from the He intershell to be burned into $^{23}$Na in the envelope. 

Table \ref{tab:na23} also illustrates that the yields of $^{23}$Na suffer from large uncertainties, up to 
two orders of magnitudes. While the upper range uncertainties are only due to the large uncertainty of the 
$^{22}$Ne($p,\gamma)^{23}$Na reaction rate, the lower range uncertainties are also determined by the effect 
of uncertainties in the $^{23}$Na$+p$ reaction rates. 
Neon-22 is an important product of IM-AGB stars, with yields of the order of $10^{-3}$ \msun\ 
in all our models (see Table \ref{tab:control}). Table~\ref{tab:ne22} shows that these numbers are very much 
affected by the large uncertainties associated with the $^{22}$Ne($p,\gamma)^{23}$Na reaction rate, as 
yields down to 0.14 of the control values 
are within the uncertainties. 
The large initial abundance of $^{20}$Ne is almost unchanged during the AGB evolution and galactic production of 
this $\alpha$-nucleus is dominated by supernova nucleosynthesis. A small but interesting range of 
variation, up to a factor of $\simeq$ 6, affects the yield of this isotope (Table~\ref{tab:ne20}). This 
uncertainty is mostly due, again, to the effect 
of the large uncertainty of the $^{22}$Ne($p,\gamma)^{23}$Na reaction combined with the feedback from the 
$^{23}$Na($p,\alpha)^{20}$Ne rate. 
We note that the NACRE rate for the 
$^{22}$Ne($p,\gamma)^{23}$Na reaction is about three orders of magnitude higher than the current 
recommended value in the HBB temperature range. Thus, model calculations performed using the NACRE rate 
would roughly correspond to taking our lower limit for the $^{22}$Ne and upper limit for the $^{23}$Na and 
$^{20}$Ne yields.

The same rate also affects the yield of $^{24}$Mg, with yield variations up to a factor $\simeq$ 5. 
However, for this isotope much larger multiplication factors, up to 48, only appear when all rates are 
simultaneously changed, and cannot be obtained by simply multiplying the uncertainties produced by each 
rate. This is because of the combined effect of the uncertainties of two reactions, 
$^{22}$Ne($p,\gamma)^{23}$Na and $^{23}$Na($p,\gamma)^{24}$Mg. Taking the upper 
limits for these two rates we find, for the 5 \msun, $Z=0.0001$ model, a 
multiplication factor of 38, close to that obtained by varying all the rates. 

The most uncertain isotope produced by the MgAl chain is $^{26}$Al. The yield of this 
radioactive isotope is very high in some of our models, reaching $\simeq 10^{-5}$ \msun. However, the large 
uncertainty associated with the rate of its destruction 
reaction, $^{26}$Al($p,\gamma)^{27}$Si, makes it possible also to have $no$ production of this isotope in 
any of our models. The very large upper limit for this rate should be tested against observational 
constraints, such as the $^{26}$Al/$^{27}$Al ratio obtained in presolar silicon carbide and oxide 
grains. It is interesting to note that the range of uncertainty of the upper limit of
$^{25}$Mg($p,\gamma)^{26}$Al reaction rate is relatively small but is almost completely 
reflected in the corresponding variations of the $^{26}$Al 
yield. In summary, the $^{26}$Al yield can be multiplied by factors down to 0.02, but also 
multiplied by factors up to 3. The yield of $^{25}$Mg itself, instead, does not feel the uncertainty 
of the $^{25}$Mg($p,\gamma)^{26}$Al reaction rate as strongly, but rather varies significantly, up to a 
factor of 1.7, only when 
all the rates are simultaneously changed. Finally, $^{26}$Mg mostly feels the uncertainties in the 
$^{26}$Mg($p,\gamma)^{27}$Al rate, while $^{27}$Al is affected by the uncertainties connected to 
both the $^{26}$Mg($p,\gamma)^{27}$Al and $^{26}$Al($p,\gamma)^{27}$Si rates, with multiplication 
factors ranging up to 6 and down to 0.39. 

\subsection{Interdependencies of the yields}

To provide uncertainty ranges as given in Tables \ref{tab:ne20} to \ref{tab:al27} is not enough if one 
wants to test these errors, for example by including them in GCE 
models. In fact, the yield variations are correlated to each other so that it is not 
possible to freely pick any combination of upper and/or lower limits for each isotope. Instead, to produce 
consistent model predictions one should test, for example, the upper limit of the $^{22}$Ne yield together 
with the lower limit of the $^{23}$Na yield and so on. In Table 13, available in electronic form
at the CDS (see footnote $\star$), we provide complete results for a relatively small subset of 
models 
obtained by varying 
only the six most uncertain reactions between their upper and lower limits. Moreover, we have 
prepared a web-interface program located at {\it http://www.astro.uu.nl/$\sim$izzard/cgi-bin/varyrates.cgi} 
through which it is possible to perform calculations for any synthetic stellar model presented in this 
paper changing any of the rates.

\subsection{Yield probability distribution}
\label{sec:examples}

\begin{figure*}
\begin{tabular}{cc}
\includegraphics[angle=270,width=8cm]{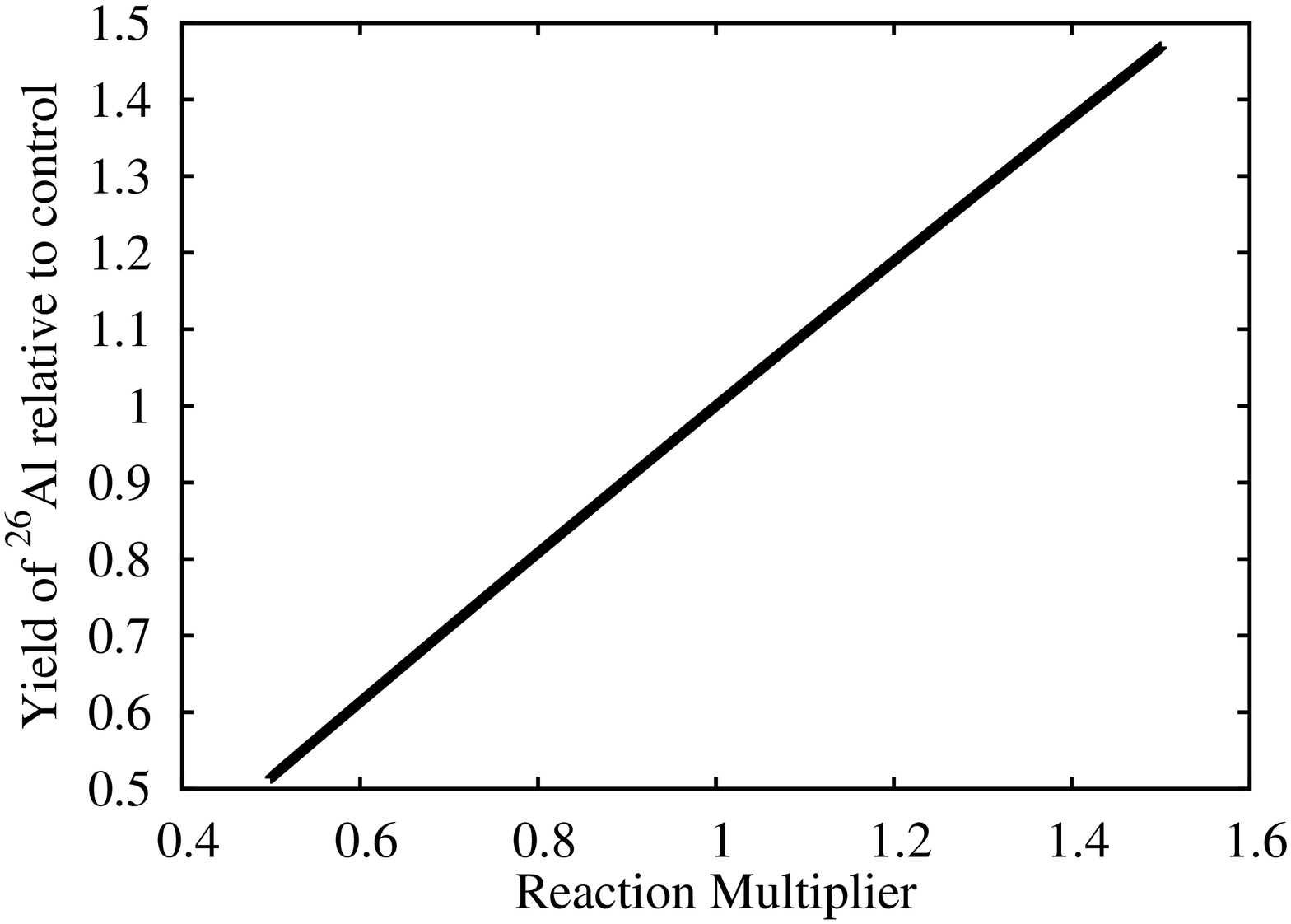} & 
\includegraphics[angle=270,width=8cm]{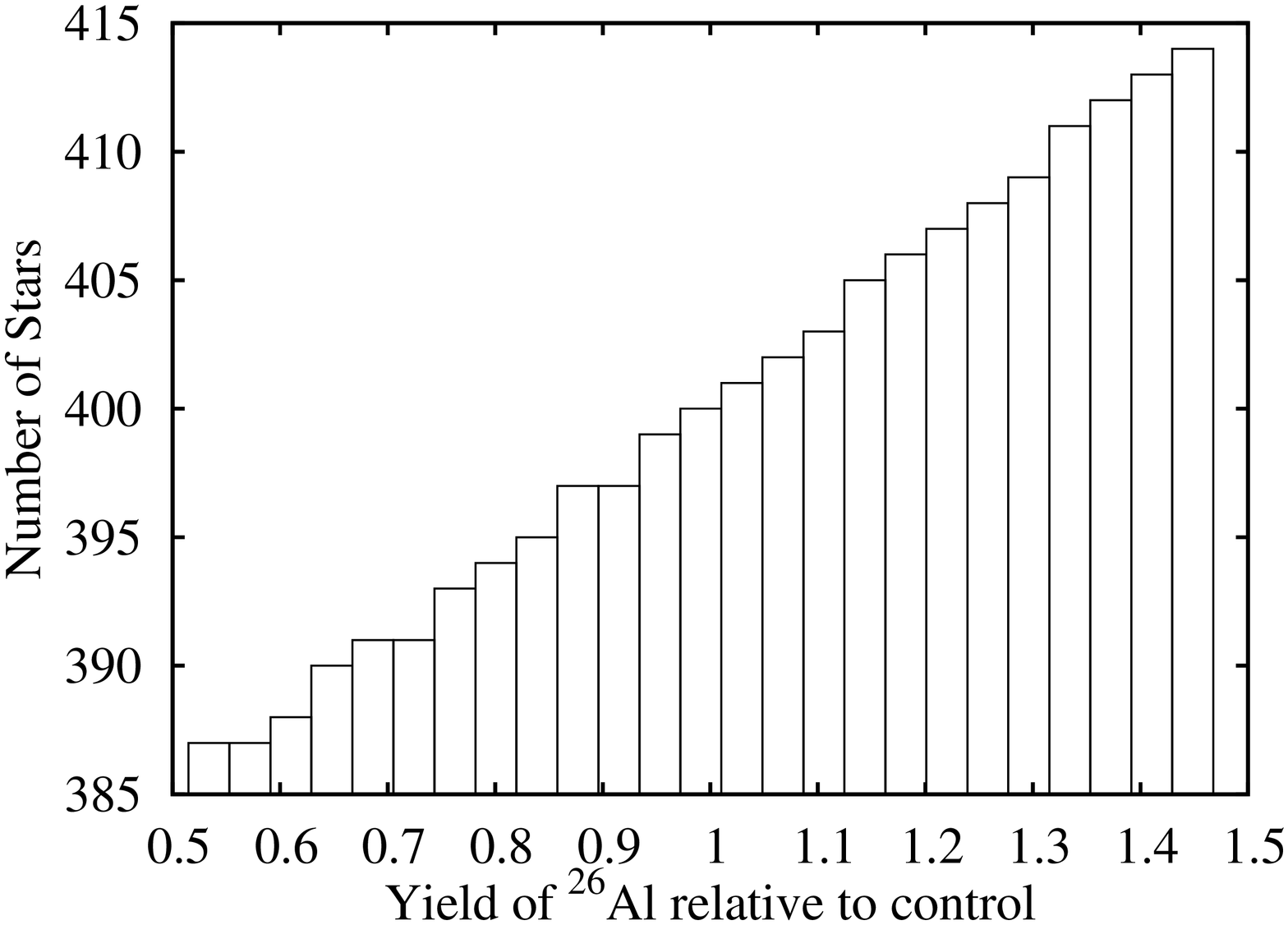} \\
\end{tabular}
\caption{For the 5 \msun\ $Z=0.008$ model we show {\em (left)}: the multiplication factor for $^{26}$Al as 
function of the variation factor of the $^{25}$Mg($p,\gamma$)$^{26}$Al reaction rate, {\em (right)}: the number of 
models obtained for each bin representing a yield interval. \label{fig:al26}}
\end{figure*}

%

\begin{figure*}
\begin{tabular}{cc}
\includegraphics[angle=270,width=8cm]{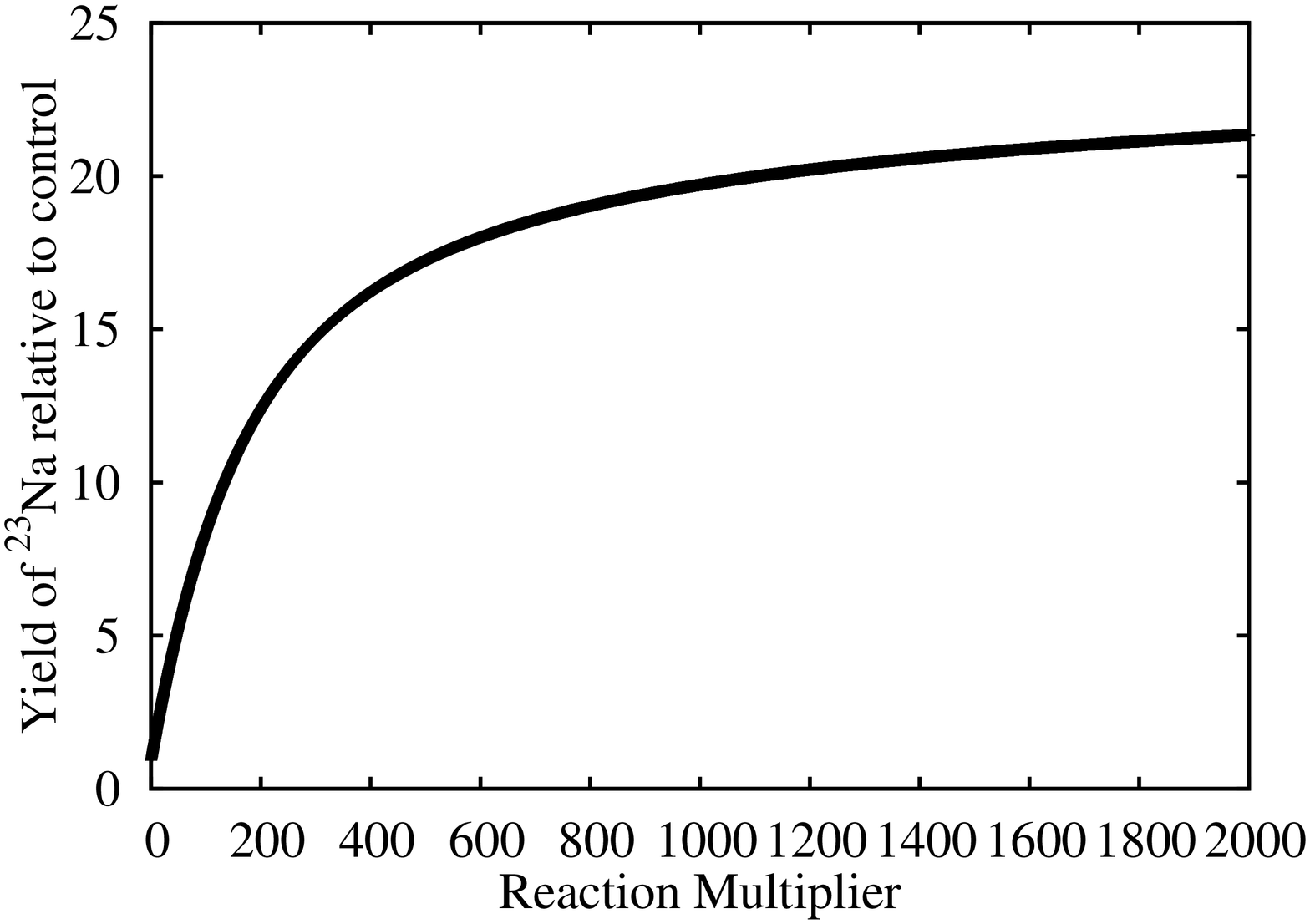} & 
\includegraphics[angle=270,width=8cm]{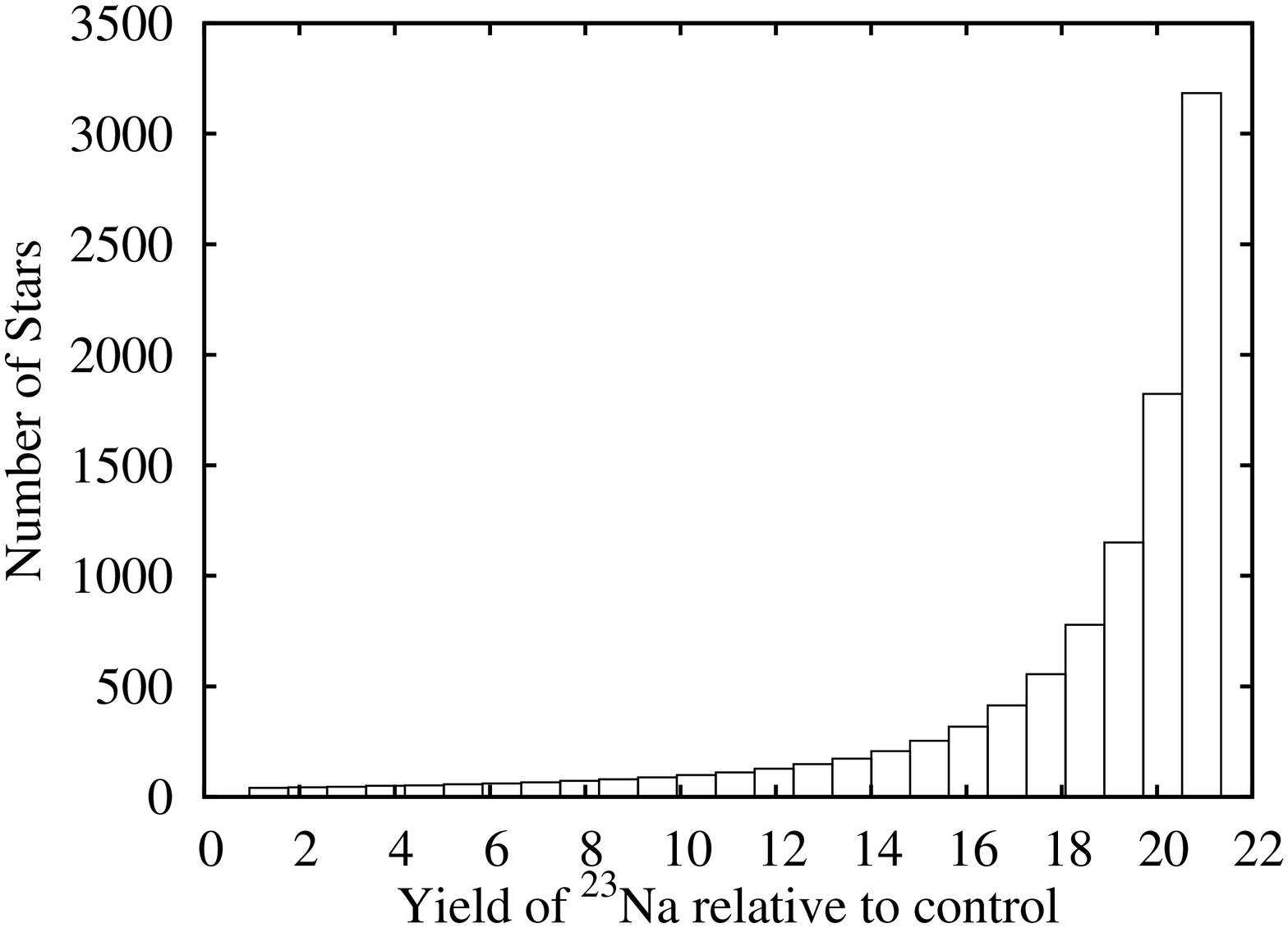} \\
\end{tabular}
\caption{For the 5 \msun\ $Z=0.008$ model we show {\em (left)}: the multiplication factor for $^{23}$Na as 
function of the variation factor of the $^{22}$Ne($p,\gamma$)$^{23}$Na reaction rate, {\em (right)}: the number of 
models obtained for each bin representing a yield interval. \label{fig:na23}}
\end{figure*}

%

As discussed in Sec.~\ref{sec:rates}, due to the difficulty of attributing a statistical significance to 
the estimates of upper and lower limits, we employed a flat probability distribution for the rates. In this section 
we show that the resulting yield distributions do not necessarily follow 
the rate distribution, i.e. they are not always flat. The following examples should only be considered as 
test exercises, and we report them in order to illustrate the future potential of our method in estimating 
yield distributions, and hence recommended values and uncertainties for the yields.  

We show two examples calculated for the 5 \msun\ $Z=0.008$ model. The first yield distribution 
is that which we obtain for $^{26}$Al when varying the $^{25}$Mg($p,\gamma$)$^{26}$Al reaction 
rate. As shown in Fig.~\ref{fig:al26} (left panel) this is a simple situation in which the 
variation in the $^{26}$Al yield is linearly dependent on the variation of the 
$^{25}$Mg($p,\gamma$)$^{26}$Al reaction rate. The resulting yield distribution is shown in 
Fig.~\ref{fig:al26} (right panel) and is almost flat, with the same number of stars (within 
5\%) reproducing the different yield ranges, represented by the bins. A more interesting 
example concerns the variation of $^{23}$Na as function of the $^{22}$Ne($p,\gamma$)$^{23}$Na 
reaction rate. The variation of the yield in this case is $not$ linearly dependent on the 
variation of the $^{22}$Ne($p,\gamma$)$^{23}$Na reaction rate (Fig.~\ref{fig:na23}, left 
panel). When the rate is slow, the reaction approaches equilibrium and the yield of $^{23}$Na 
increases linearly with the rate. As the rate becomes faster, the yield of $^{23}$Na stops its 
rapid rise because almost all the $^{22}$Ne fuel is used up. The $^{23}$Na/$^{22}$Ne ratio is 
nearly at its equilibrium value, but the amount of extra mass converted to $^{23}$Na as a 
result of the increased rate is small because there is simply no more $^{22}$Ne fuel. This is 
complicated because the third dredge up replenishes the supply of $^{22}$Ne and allows some 
extra $^{23}$Na to be produced. This leads to the shallow slope at high rates. Note that 
$^{22}$Ne is always produced during the final few pulses when the temperature at the base of 
the convective envelope falls below that required to activate the 
$^{22}$Ne($p,\gamma$)$^{23}$Na reaction. Consequently, the $^{23}$Na yield distribution leans 
towards high values, around $\simeq$20 of the multiplication factor, and these values are more 
likely to occur than smaller values (Fig.~\ref{fig:na23}, right panel). This is a consequence 
of the fact that we have chosen a flat distribution to represent the rate uncertainties. We 
stress again that this choice is still quite arbitrary, and these yield distributions are test 
exercises. We cannot give recommendations on the most likely value or the standard deviation 
from the yield distributions and we suggest, for the time being, all the values that we obtain 
for each yields to be equally probable.

The left panel of Fig.~\ref{fig:na23} is also useful to demonstrate that our choice of a constant value of 
2000, instead of the actual range of 982 to 1888, for the upper limit factor of the 
$^{22}$Ne($p,\gamma$)$^{23}$Na rate does not affect the accuracy of the results, since the results do not 
change once the rate is multiplied by more than 800. 
For the $^{26}$Al$(p,\gamma$)$^{27}$Si rate our results show that, similarly to the case of $^{23}$Na discussed 
above, the effect of varying this rate are more or 
less the same once the rate has been multiplied by a factor $\simeq$ 200 because for this value of the rate all 
$^{26}$Al is consumed and the yield distribution is weighed towards the low $^{26}$Al yields. However, contrarily to 
the case of the $^{22}$Ne($p,\gamma$)$^{23}$Na rate discussed above, the upper limit of the 
$^{26}$Al$(p,\gamma$)$^{27}$Si rate ranges 
below this saturation value, hence we should take the results obtained changing this rate as maximum possible 
ranges. We have tested this point by running the detailed $M=6$ \msun\ $Z=0.02$ and $Z=0.004$ models 
using the proper upper 
limit for the $^{26}$Al$(p,\gamma$)$^{27}$Si rate, i.e. including an accurate description of its variation with the 
temperature and found that in fact the ranges of variation are smaller than those reported in Table~\ref{tab:al26}: 
0.67 rather than 0.43 for the $Z=0.02$ model, and 0.19 rather than 0.02 for the $Z=0.004$ model. For $^{27}$Al, 
instead, the ranges of variation are unchanged: 1.47 rather than 1.59 for the $Z=0.004$ model. The 
same point holds for the lower limit of this rate. 
The same problem arise for the $^{26}$Mg$(p,\gamma$)$^{27}$Al reaction, whose 
uncertainty factors (Table~\ref{tab:rates}) also vary significantly with the temperature. Thus, we should 
consider also the yield variations obtained varying this rate as the maximum allowed. 
For the $^{23}$Na($p,\gamma$)$^{24}$Mg 
reaction, the lower limit of the rate show a large range of variations, however, our results show that the 
uncertainty associated with the lower limit of this rate does not affect any isotope to more than 20\%. 
Finally, for the $^{27}$Al$(p,\gamma$)$^{28}$Si reaction our 
results have showed that changes in this rate do not affect any isotopic yield. 
This result was confirmed by a detailed $M=6$ \msun\ $Z=0.02$ model computed using the upper 
limit of the rate. In summary, only the results 
obtained by varying the $^{26}$Mg$(p,\gamma$)$^{27}$Al and the $^{26}$Al$(p,\gamma$)$^{27}$Si reactions 
should be taken as the maximum allowed.

\section{Summary and conclusions}

We have shown that uncertainties in the yields of the Ne, Na, Mg and Al isotopes 
are present in connection to proton-capture reaction rates. The most uncertain rates are those of  
the $^{26}$Al($p,\gamma$)$^{27}$Si and the $^{22}$Ne($p,\gamma$)$^{23}$Na reactions, with variations up to 3 
orders of magnitude. The 
$^{22}$Ne($p,\gamma$)$^{23}$Na uncertainties produce huge variations, up to two orders of magnitude, in the 
yields of $^{22}$Ne and $^{23}$Na, as well as uncertainties up to a factor of 5 in the yields of $^{20}$Ne and 
$^{24}$Mg. 
The $^{26}$Al($p,\gamma$)$^{27}$Si uncertainties lead to large variations, up to two orders of magnitude, 
in the
yields of $^{26}$Al. The yield of $^{24}$Mg is also 
affected by the uncertainty in the $^{23}$Na($p,\gamma$)$^{24}$Mg rate, with strong effects appearing when 
this rate is considered together with the $^{22}$Ne($p,\gamma$)$^{23}$Na rate. 
The effect of the relatively small uncertainty of the $^{25}$Mg($p,\gamma$)$^{26}$Al reaction rate turned out to  
be quite important because of being completely, and not only partially as for the other 
rates, reflected in the uncertainty of the $^{26}$Al yields. Finally, the $^{26}$Mg($p,\gamma$)$^{27}$Al 
affects the yields 
of both $^{26}$Mg and $^{27}$Al. In summary, in order to obtain more reliable Ne, Na, Mg and Al yields from 
IM-AGB stars the rates to be better determined are: 
$^{22}$Ne($p,\gamma$)$^{23}$Na, $^{23}$Na($p,\gamma$)$^{24}$Mg, 
$^{25}$Mg($p,\gamma$)$^{26}$Al, $^{26}$Mg($p,\gamma$)$^{27}$Al, and $^{26}$Al($p,\gamma$)$^{27}$Si. 
 
It is difficult to predict {\it a priori} exactly what will be the impact of our uncertainty ranges and 
detailed models should be constructed to address each of the 
observational constraints related to HBB nucleosynthesis. With regards to globular cluster anomalies, 
the Na abundances predicted by Fenner et al. (\cite{fenner:04}) to be too large to match the observed abundances 
correspond to the upper limits of the Na yields and a revision of this point is necessary. On the other hand, the 
predicted too high Mg abundances may be more difficult a problem to solve taking into account reaction rate 
uncertainties, while the observed high Al abundances may also be matched within uncertainties.  
Only performing detailed calculations will it be possible to verify if a solution for globular cluster 
anomalies is feasible within the uncertainties we have presented here. 

\begin{acknowledgements}
RGI and ML (VENI fellow) are supported by the NWO.
Part of the computations were performed on CITA's Mckenzie cluster which was funded
by the Canada Foundation for Innovation and the Ontario Innovation Trust.
\end{acknowledgements}

\end{document}